\documentclass[epj]{svjour}
\usepackage{graphics}
\newcommand{\be}{\begin{equation}}
\newcommand{\ee}{\end{equation}}
\newcommand{\bea}{\begin{eqnarray}}
\newcommand{\eea}{\end{eqnarray}}
\newcommand{\no}{\noindent}

\newcommand{\e}{{\rm e}}
\newcommand{\bit}{\begin{itemize}}
\newcommand{\eit}{\end{itemize}}
\newcommand{\bico}[2]{
{\Big(\!\begin{array}{c}{#1}\\{#2}\end{array}\!\Big)}}
\begin{document}
\title{Collective Effects From 
Induced Behaviour}
\author{Ion-Olimpiu Stamatescu\inst{1}\inst{2} 
\and Tai Tsun Wu\inst{3}\inst{4}
\thanks{Work supported in part by the United States Department of Energy 
  under Grant No. DE-FG02-84ER40158.}%
}                     
%
%
\institute{FESt, Schmeilweg 5, D-69118 Heidelberg, Germany
 \and Institut f\"ur Theoretische Physik, Philosophenweg 16,
D-69120 Heidelberg, Germany\and Gordon McKay Laboratory, Harvard University, Cambridge, MA 02138, USA\and Theory Division, CERN, CH-1211 Geneva 23, 
Switzerland}
\date{Received: date / Revised version: date}
%
\abstract{
 We present a solvable model for
describing quantitatively situations where the individual behaviour 
of agents in a group ``percolates" to collective  behaviour 
of the group as a whole as a result of mutual influence between the 
agents.
} 
\maketitle
\section{Introduction}
\label{s.intro}
One very common interaction between the individuals of a group
is the tendency to imitate each other. Under
certain conditions this might be expected to critically affect the
behaviour of the group as a whole.  It may be
interesting to have a quantitative understanding of the conditions
under which individual patterns of behaviour may
 propagate to a group behaviour
as a result of mutual influence among a few neighbors.
As another example we may consider the problem of ensuring the reliability in
 functioning of a complex machinery by increasing  redundancies. Typically
 the ``parallel" components cannot be made completely
independent.
How does reliability generally depend on
redundancy if failure propagation is possible? This is one major
question in security analyses \cite{rasm}.
There are further examples with similar
problem setting: the dynamics of certain phase transitions, critical reactions,
epidemic models, etc.

Here we would like to analyze how {\it induced behaviour} 
can lead to
{\it collective effects} in the frame of a probabilistic model 
introduced earlier \cite{rnt} and which can 
be solved exactly. Thereby we restrict ourselves to the very 
elementary mechanism of imitation  and we do not attempt to 
include more refined interactions, such as beliefs, goals, cooperation, 
competition, etc
 -- see, e.g., \cite{HG}. For this case we can 
provide a solution in closed form. 
The model will be described and discussed 
in section 2
and its  solution in section 3. Here we also define 
a Monte-Carlo simulation by interpreting the
closed solution as a partition function. 
Approximations permiting some qualitative insight are discussed in Section 4.
In Section 5 we 
present and discuss the results for some representative cases. Here we 
use the Monte-Carlo simulation  to treat
 large ensembles of agents with ``realistic" correlations.

\section{The probabilistic model}
\label{s.model}

We consider a set ${\cal N}$ consisting of 
$n$ points, labeled $i = 1, \dots , n$, each of which can
{\it spontaneously} burst (and ``disappear") with probability $w_{0i}$ and 
let
 $K_{ij}$ be the {\it induced} probability that point $i$ 
bursts 
{\it because} $j$ {\it has} bursted.

If these points represent the parallel components of a machinery, the
functioning of the latter is ensured as long as at least one of the 
components
works. For the behaviour of a group
of agents the relevant question is again whether (practically) all
agents show the same behaviour. 
The key quantity is therefore the probability with which all points have 
bursted: 
\be
W = W(n) = W(n;\{w_0,K\}).
\ee
\no We also define
the ``no-propagation" probabilities
\be
L_{ij}=1-K_{ij}
\label{e.l}
\ee
\no and introduce the following
simplifying assumptions:

\bit
\item [1)] Symmetry:
\label{e.sym}
\be
K_{ij} = K_{ji}; 
\ee

\item [2)] Independence of ``no-propagation" events: 
\be
L_{1(23)} = L_{12}L_{13},
\ee
\no i.e. 
the probability that point 1 bursts because 2 and 3 
have both bursted is
\be
K_{1(23)} = 1 - (1-K_{12})(1-K_{13}) =K_{12}+K_{13}-K_{12}K_{13}.
\label{e.ind}
\ee
\eit

Notice that the absence of time evolution in this model implies
 that, in a real situation, the whole
development is expected to 
take place in a very short time such that no parameter
changes appreciably.
The simplifying assumptions 1), 2)  are of course limitations, 
we can imagine, however,
many realistic situations under which they hold approximately. 
Symmetry, for instance, 
may well be expected to hold on the average in a group of similar 
agents (the birds in a flock, say). The independence assumption 
 depends
 on the real interactions. 

For illustration consider the following 
``temperature model":
3 points isolated in an enclosure, with the bursting probability for
point `1' described by some monotonic function of the ambient temperature. 
Assume that bursting of a point increases the 
average temperature by $\Delta T$.  We ignite points `2' and `3'  and see what 
happens
with `1'. If only `2' {\it or} `3' had bursted, `1' will go off with probability
$P(\Delta T)$, while if both `2' {\it and} `3' went off the ambient temperature 
is 
$2\Delta T$ and `1' will
explode with probability $P(2\Delta T)$. Then (\ref{e.ind}) would require:
\be
P(2\Delta T) = 2P(\Delta T) -P(\Delta T)^2,
\label{e.tem1}
\ee
\no or, with $h(T) = -ln(1-P(T))$,
\be
h(2T)= 2h(T),
\label{e.tem2}
\ee
\no with solution $h(T)=aT$. Hence in this ``temperature model" only:
\be
P(T)=1-\e^{-aT}
\ee
\no is compatible with (\ref{e.ind}). In particular, a threshold behaviour
like $P(T) = \theta(T-T_0)$ will violate (\ref{e.tem1}) if
$\Delta T < T_0 < 2 \Delta T$. We shall retain, however,
for this analysis the
independence assumption (we shall indicate below how one can relax it
when the dynamics of the interaction is known).

Before we proceed and solve the model notice the following four limiting cases:
\bea
 K_{ij}=0&:&\ \ W = W_0(n) =\prod_{i=1}^n w_{0i}; \label{e.l0} \\
 K_{ij}=1&:&\ \ W = W_1(n) =1 - \prod_{i=1}^n (1- w_{0i}); \label{e.l1}\\
 w_{0i}=0&:&\ \ W=0 \label{e.w0}; \\
 w_{0i}=1&:&\ \ W=1 \label{e.w1}.
\eea
\no They are helpful for tests, normalization, etc.

\section{Exact solutions}
\label{s.sol}
\subsection{Combinatorial solution.}
\label{s.comb}

Starting to solve the model we first remark:\par\medskip   

\no {\it in following the propagation of the bursts each $K_{ij}$ 
can be used only once: 
there is no way to get something like $K^2_{ij}$.}\par\medskip

\no It is convenient to introduce the notation $\alpha =(ij)$ for the 
non-ordered pair $\{i,j\}$. There are $\frac{1}{2}n(n-1)$ different
$\alpha$'s, $\{\alpha \} = \Omega$. 
Let $\omega$ denote a subset of $\alpha$'s; there are 
$2^{\frac{1}{2}n(n-1)}$ different $\omega$'s (including the empty set
$\emptyset$).
For the time being we shall take $w_{0i}=w_0$ independent on $i$, 
then $W$ is of the form:
\be
W = W(n,w_0,\{K_{\alpha}\}) = \sum_{\omega}C_{\omega}
\prod_{\alpha \in \omega}K_{\alpha},
\label{e.wck}
\ee
\no with the last factor taken to be 1 for $\omega = \emptyset$. 
Putting $K=1$ on a subset of $\Omega$ and 0 in the rest
we define:
\be
W_{\omega} = W |_{K_{\alpha}=1 \ \hbox{ if } \alpha \in \omega,\ 
K_{\alpha}=0 \ \hbox{ if } \alpha {\not{\in}} \omega} 
\label{e.wo}
\ee
\no and we have:
\be
W_{\omega} = \sum_{\omega' \subseteq \omega}C_{\omega'}
\prod_{\alpha \in \omega'}K_{\alpha}|_{K_{\alpha}=1} =
\sum_{\omega' \subseteq \omega}C_{\omega'}.
\label{e.wc}
\ee
\no Let for $\omega' \subset \omega$:
\be
\pm _{\omega \omega'} = (-1)^{(\hbox{nr. of elements of } \omega)
- (\hbox{nr. of elements of } \omega')},
\label{e.pm}
\ee
\no then we can invert (\ref{e.wc}) to obtain
\be
C_{\omega}=\sum_{\omega' \subseteq \omega}
\pm_{\omega \omega'}W_{\omega'}.
\label{e.cw}
\ee

We next evaluate $W_{\omega}$. Each $\omega$ achieves a 
partition 
of ${\cal N}$ in the following way: if $\{i,j\} \in \omega$ we 
join the points $i$ and $j$. Thus  ${\cal N}$ is 
partitioned into (non-empty) connected subsets $\nu_k^{(\omega)}$ 
labeled with the index $k$ and containing 
$n_1^{(\omega)},n_2^{(\omega)},
\dots$ points, $n_k^{(\omega)}>0$, such that $\sum_k n_k^{(\omega)} = n$. 
Then:
\be
W_{\omega} = \prod_k \left(1-(1-w_0)^{n_k^{(\omega)}}\right)
\label{e.ww1}
\ee
\no and thus from (\ref{e.cw}), (\ref{e.wck})
\be
W = \sum_{\omega \subseteq \Omega}\sum_{\omega' \subseteq \omega}
\pm _{\omega \omega'}
\prod_k \left(1-(1-w_0)^{n_k^{(\omega')}}\right)\prod_{\alpha \in 
\omega}K_{\alpha}.
\label{e.wpm}
\ee
\no After rearranging the terms using (\ref{e.pm}), (\ref{e.wpm}) gives:
\bea
W(n; w_0,\{K\}) &=& \sum_{\omega \subseteq \Omega}\prod_k\left(1- 
(1-w_{0})^{n_k^{(\omega)}}\right) \times \nonumber \\
& &\prod_{\alpha \in 
\omega}K_{\alpha}\prod_{\alpha' \in C^{\Omega}_{\omega}}(1-K_{\alpha'}).
\label{e.ww}
\eea
\no The extension to different $w_{0i}$ is straightforward and leads to
the general solution:
\bea
W(n;\{w_0,K\}) &=& \sum_{\omega \subseteq \Omega}
\prod_k \left(1-\prod_{i \in \nu_k^{(\omega)}}(1-w_{0i})\right)
\times \nonumber \\
& &\prod_{\alpha \in 
\omega}K_{\alpha}\prod_{\alpha' \in C^{\Omega}_{\omega}}(1-K_{\alpha'}).
\label{e.wt}
\eea
\no The sum over $\omega$ is taken over all subsets of
$\Omega $, including the empty set and $\Omega$, where $\Omega$ is the
set of all $\alpha$'s, and $C^{\Omega}_{\omega}$ is the complement of the 
set $\omega$ in ${\Omega}$.\par\bigskip

\subsection{Iterative solution.}
\label{s.iter}
 
It is helpful to write down also an iterative solution of the model. Consider
a set $\eta$ of points out of which only a subset $\sigma$ is still 
untouched, and consider all the ways the failure can propagate from the 
points in $C^{\eta}_{\sigma}$ to those in $\sigma$, then the probability
that also $\sigma$ blows up is:
\bea
&&P(\eta|\sigma) = \sum_{\tau \subset \sigma} \prod_{i \in C_{\tau}^{\sigma}}
\left(1-\prod_{j \in C_{\sigma}^{\eta}}L_{ij}\right) \times \nonumber \\
&&\prod_{i \in \tau}
\left(\prod_{j \in C_{\sigma}^{\eta}}L_{ij}\right) P(\sigma|\tau), \ 
\ \ P(\eta|\emptyset) = 1
\label{e.wit1}
\eea
\no and we have:
\be
W(n;\{w_0,K\}) = 
\sum_{\sigma \subset {\cal N}} \prod_{i \in C^{\cal N}_{\sigma}}
w_{0i} \prod_{j \in \sigma} (1-w_{0j}) P({\cal N}|\sigma). 
\label{e.wit2}
\ee
\no In (\ref{e.wit1},\ref{e.wit2}) all inclusions are strict and go also over 
the empty set.

For the case of ``homogeneous interaction" 
\be
K_{ij}=K, \ \ L_{ij} = L,
\label{e.kitc}
\ee
\no (\ref{e.wit1},\ref{e.wit2}) simplify
considerably:

\bea
P_{m,l} = \sum_{k=0}^{l-1} \bico{l}{k}
(1-L^{m-l})^{l-k}(L^{m-l})^l
P_{l,k}  \label{e.witc1} \\
W(n;w_0,K) = \sum_{m=0}^{n-1} \bico{n}{m}w_0^{n-m}(1-w_0)^mP_{n,m}.
\label{e.witc2}
\eea 

Finally let us remark that in the frame of an explicit model for the 
interaction one can relax the assumption (\ref{e.ind}) and directly 
construct the compound probabilities appearing in (\ref{e.wit1}).
For instance, for the ``temperature model" of section 2 we only need 
to substitute in (\ref{e.wit1}):
\be  
\prod_{j \in C_{\sigma}^{\eta}}L_{ij} \rightarrow 
1-P_i\left(\sum_{j \in C_{\sigma}^{\eta}}T_{ij}\right),
\label{e.witemp}
\ee
\no where $T_{ij}$ is the increase in temperature at site `i' due to 
the bursting of point `j' and $P_i(T)$ is the probability that point `i' 
explodes when the ambient temperature is $T$.

\subsection{Monte Carlo analysis}
\label{s.mc}

A Monte Carlo simulation can be set up  based on (\ref{e.ww},\ref{e.wt}). 
We define for arbitrary $p$ partition functions:
\be
{\cal Z}_p = \sum_{\omega \subseteq \Omega}
{\cal W}_0(\omega)^p
\prod_{\alpha \in
\omega}K_{\alpha}\prod_{\alpha' \in C^{\Omega}_{\omega}}(1-K_{\alpha'}),
\label{e.mc1}
\ee
\no where
\be
{\cal W}_0(\omega) =  \prod_k\left(1-
(1-w_{0})^{n_k^{(\omega)}}\right)
\label{e.mc2}
\ee
\no and we have  (see (\ref{e.w1})):
\be
{\cal Z}_0 = 1,\ \  {\cal Z}_1= W(n).
\label{e.mc22}
\ee
\no  Starting from
any partition function ${\cal Z}_p$ we can write $W(n)$ as an average:
\bea
&&W(n) = {{\langle {\cal W}_0^{1-p} \rangle_{p}} \over
         {\langle {\cal W}_0^{-p} \rangle_{p}}},\\ 
&&\langle {\cal W}_0^{q} \rangle_{p}= 
\frac{1}{{\cal Z}_p}
\sum_{\omega \subseteq \Omega}{\cal W}_0(\omega)^{q}{\cal W}_0(\omega)^{p}
\prod_{\alpha \in
\omega}K_{\alpha}\prod_{\alpha' \in C^{\Omega}_{\omega}}(1-K_{\alpha'}),
\nonumber \label{e.mc3}
\eea

\no in particular (see (\ref{e.mc22})):
\be
W(n) = \langle {\cal W}_0 \rangle_{0} .
\label{e.mc0}
\ee

The MC procedure uses the terms in ${\cal Z}_p$ as Boltzmann-Gibbs
factors to achieve an importance sampling of $\nu_{k}^{\omega}$ partitions.
In actual simulations using $p=0$, 0.5 or 1 the results were similar, 
therefore we used for the 
systematic analysis
 $p=0$, i.e. eq. (\ref{e.mc0}), which is faster. 
Then the Metropolis algorithm, which produces new partitions 
by adding or removing ``bonds" $\alpha =(ij)$
one at a time, is local (and vectorizable). Note that since the
${\cal W}_0$ are positive, lack of convergence in the MC simulation
based on (\ref{e.mc0}) is likely to show up as underestimation of the exact 
result.

Whenever
we could compare the results of 
the Monte Carlo simulation with exact summation of either
the combinatorial (\ref{e.ww},\ref{e.wt}) 
or the iterative (\ref{e.wit1},\ref{e.wit2}) solution
 we have found very good agreement -- see section 5.

\section{Approximations}
\label{s.appr}

\subsection{Mean Field approximation}
\label{s.mf}

We
introduce an ``effective" bursting probability $w_i$ via the consistency
equation
\be
1-w_i = (1-w_{0i})\prod_{j \neq i}(1-w_jK_{ij}).
\label{e.mfc}
\ee
\no We shall in the following assume translational invariance, that is, $w_{0i}=w_0$
and $K_{ij}=K_{|i-j|}$. For finite systems we shall assume periodic boundary 
conditions. Then the mean field equation reads:
\be
1-w=(1-w_0){\rm exp}\left( \sum_{\nu=1}^{n-1} {\rm ln}(1-wK_{\nu})\right). 
\label{e.mfe}\ee
\no  $w$ can be found iteratively. A rough estimate is:
\be
w \simeq 1 - (1-w_0){\rm e}^{-\lambda{\cal K}} , \nonumber \ \ \
{\cal K} = \sum_{\nu=1}^{n-1} K_{\nu}
\label{e.kal}
\ee
\no with some $\lambda \sim {\cal O}(1)$. The parameter  ${\cal K}$
has an intuitive meaning: it gives the average number of points 
which can be affected by one point. We then have:
\be
\hbox{ln}W(n) = n{\rm ln}w \sim -(1-w_0)\hbox{e}^{{\rm ln}n - \lambda {\cal K}}, 
\label{e.Wmean}
\ee
\no which indicates that the behaviour of $W(n)$ is determined by
the dependence on $n$ of ${\cal K}$.

\subsection{A lower limit}
\label{s.aap}

A more refined approximation can be derived which, 
for the case of homogeneous interaction (\ref{e.kitc}) provides a 
lower bound for $W(n)$. We start from 
the iterative solution (\ref{e.wit1},\ref{e.wit2}) and assume that for
 some $\lambda$ we have:
\bea
P(\sigma|\tau) \geq \prod_{i \in \tau} (1-F_i(\sigma ;\lambda))  \\ 
F_i(\sigma ;\lambda) \equiv \prod_{j \in \sigma,\ j \neq i}(1-
\lambda K_{ij})
\label{e.appr1}
\eea
\no  (here 
$\emptyset \subset \tau \subset \sigma \subset \eta$). Since
all contributions to (\ref{e.wit1}) are positive we can then write:
\bea
&&P(\eta|\sigma) \geq 
\sum_{\tau \subset \sigma} \left[ \prod_{i \in C_{\tau}^{\sigma}}
\left(1-\prod_{j \in C_{\sigma}^{\eta}}L_{ij}\right) \right]  \times
 \nonumber \\
&&\left[\prod_{i \in \tau} \left(\prod_{j \in C_{\sigma}^{\eta}}L_{ij}\right)\right] 
\left[
\prod_{i \in \tau} (1-F_i(\sigma ;\lambda))\right]  \nonumber \\
&&= \prod_{i \in \sigma}\left(1-F_i(\sigma ;\lambda)
\prod_{j \in C_{\sigma}^{\eta}}L_{ij}\right) - \nonumber \\
\prod_{i \in \sigma}&& \left( (1-F_i(\sigma ;\lambda))
\prod_{j \in C_{\sigma}^{\eta}}L_{ij}\right) \label{e.appr2} 
 \geq\prod_{i \in \sigma} (1-F_i(\eta ;\lambda)) .
\label{e.appr3}
\eea
\no If we can prove that there exists a $0 \leq \lambda \leq 1$
such that the last inequality holds, then we have:
\bea
W(n;\{w_0,K\}) &\geq&  \prod_{i \in {\cal N}}\left(1 -
F_i({\cal N};\lambda)(1-w_{0i})\right) - \nonumber \\
& &\prod_{i \in {\cal N}}
(1-F_i({\cal N};\lambda))(1-w_{0i}) .
\label{e.appr4}
\eea

\begin{figure}[tb]
\vspace{6.5cm}
\includegraphics{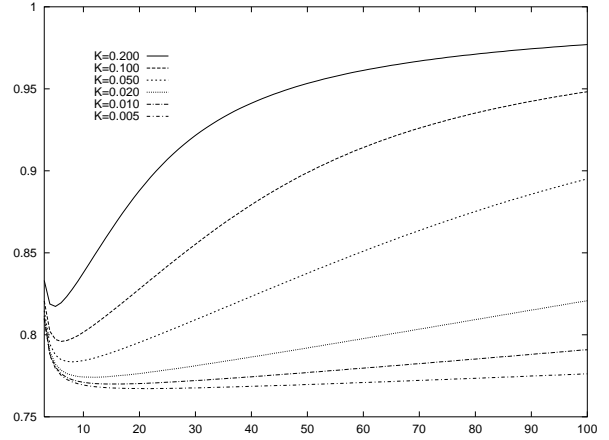}
\caption{{\it $\lambda$ satisfying (\ref{e.ine2}) vs $n$ for various $K$.
For values of $\lambda$ below the curves the formulae 
(\ref{e.appr1},\ref{e.appr4}) provide  lower bounds.
}}
\label{f.lbo}
\end{figure}

In the homogeneous interaction case we can show that 
  a $\lambda>0$ can always be found such that (\ref{e.appr3}) holds.
We have:
\be
F_i(m;\lambda) = (1-\lambda K)^{m-1}
\ee
\no and we must show that for all $m,\ l < m$:
\bea
&&\left[1-(1-\lambda K)^{l-1}(1-K)^{m-l}\right]^l 
\geq \left[1-(1-\lambda K)^{m-1}\right]^l \nonumber \\
&&+ \left[(1-(1-\lambda K)^{l-1})(1-K)^{m-l}\right]^l.
\label{e.ine1}
\eea
\no The expression
 in the square brackets on the left hand side decreases with 
increasing $l$, while those on the right hand side do not. Therefore the
worst case is  
$l=m-1$ and it is enough to prove that:
\bea
&&\left[1-(1-\lambda K)^{m-2}(1-K)\right]^{m-1} \geq
\left[1-(1-\lambda K)^{m-1}\right]^{m-1} \nonumber \\
&& + \left[(1-(1-\lambda K)^{m-2})(1-K)\right]^{m-1} .
\label{e.ine2}
\eea 
\no For $m=2$ (\ref{e.ine2}) is satisfied for any $\lambda < 1$. 
For $m \sim 1/K$ (\ref{e.ine2}) is satisfied for  $\lambda \geq 1/2$
and with increasing $m$ the bound on $\lambda$ goes toward 1. 
More precise numerical bounds are given in Fig. \ref{f.lbo}. For the 
general correlation case we may use  (\ref{e.appr4})
as an alternative to the mean field approximation.

\section{Results and Discussion}
\label{s.res}

In the introduction we asked about the collective effects of
induced behaviour of agents in an ensemble. As we have noticed, 
various specific questions can  be asked in this context.
 They can all be subsumed under some general questions, 
which in the above model can be exemplified as follows:
\par\medskip   

\no {\bf Question 1:} {\it How does the total burst (failure)
probability 
$W(n,w_0,\{K_{ij}\})$ 
behave with increasing $n$ for various types of ``aggregation", 
distinguished by the way in which
the mutual influence between systems depends on $n$? 
}\par\medskip

As an instructive example we consider a 1-dimensional 
ensemble and put $n$
points equidistantly on a circle. 
We assume ``finite correlation length" $\xi$:
\bea
K_{ij}&=& a \hbox{e}^{- d_{ij}/{\xi}}, \label{e.kij}\\
d_{ij}&=& {\rm min}(|i-j|,n-|i-j|),
\label{e.kdist}
\eea
\no with 
\be 
\xi(n) = \xi_0 n^{\alpha}.
\label{e.xi}
\ee
\no Hence for $\alpha=1$ we have an {\it intensive} aggregation 
(more and
more points come under the influence of a single one while the
size of the ensemble measured in correlation lengths stays fixed)
and for
$\alpha=0$ an {\it extensive} aggregation (the density of points stays
constant while the total volume increases). $0 < \alpha < 1$ interpolates between
these situations (we are not concerned here with the dynamics of the
aggregation: attraction, repulsion  etc but just take the aggregation law
as given).

\begin{figure}[tb]
\vspace{6.5cm}
\includegraphics{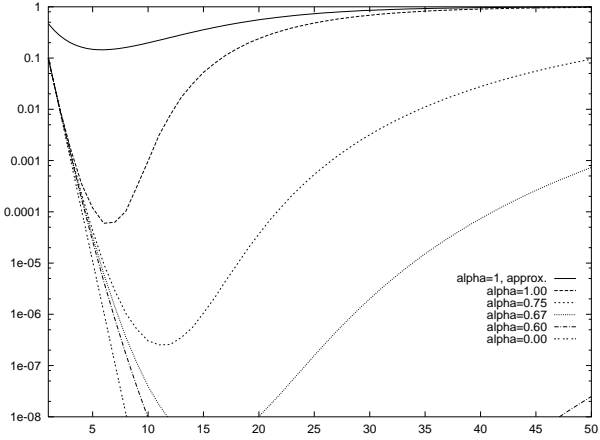}
\caption{{\it $W(n)$  vs $n$ from the mean field approximation using 
the iterative
solution of eq. (\ref{e.mfe}); the line identified by ``approx." uses
the further approximation (\ref{e.Wappro}). We use
$w_0=0.1,\ a=0.3,\ \xi_0=1/3.5$ -- see (\ref{e.kij}-\ref{e.xi}). 
The different aggregation types are 
identified by $\alpha$.
}}
\label{f.mfa}
\end{figure}

A rough impression is offered by the mean field approximation.
 For large $n$ and small
$w_0$, $K$:
\be 
{\cal K} \simeq {a}{\xi_0} n^{\alpha} \label{e.Kappro}
\ee
\no and
\be 
{\rm ln}W(n) \sim -(1-w_0)\hbox{exp}\left({\rm ln}n-
{a \lambda}{\xi_0} n^{\alpha}\right),
\label{e.Wappro}
\ee
\no see (\ref{e.mfe})-(\ref{e.Wmean}), which has a minimum  for:
\be
n_{\rm min} \sim \left({a \alpha \lambda \xi_0}
\right)^{-\frac{1}{\alpha}},
\label{e.nmin}
\ee
\no above which $W(n)$ goes to 1 for all $\alpha > 0$. In Fig. \ref{f.mfa}
we illustrate this behaviour, both from formula (\ref{e.Wappro}) and using an 
iterative solution of eq. (\ref{e.mfe}).  

\begin{figure}[tb]
\vspace{6.5cm}
\includegraphics{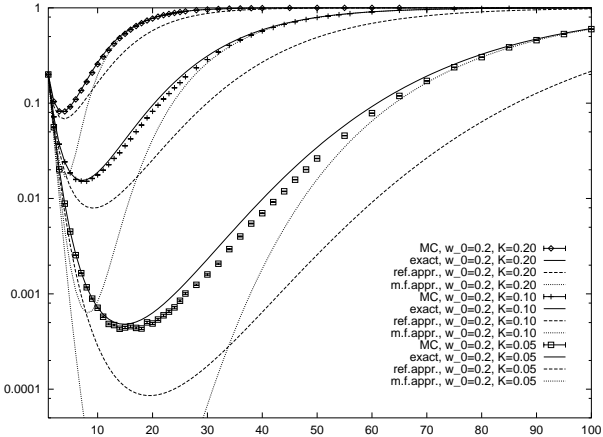}
\caption{{\it $W(n)$ for homogeneous interaction  
(infinite correlation length) from: the 
 Monte Carlo simulation, the exact summation of the iterative solution 
(\ref{e.witc1},\ref{e.witc2}), the ``refined approximation"
(\ref{e.appr4}) (with the corresponding minimal values of $\lambda$
from Fig. \ref{f.lbo})
and the mean field approximation (\ref{e.mfe}).
Here $w_0=0.2$, $K=0.05$, $0.1$ and $0.2$. 
}}
\label{f.xk0}
\end{figure}

\begin{figure}[tb]
\vspace{6.5cm}
\includegraphics{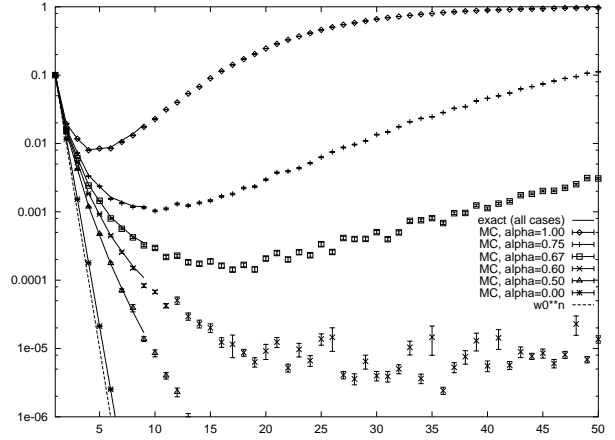}
\caption{{\it $W(n)$  vs $n$ from the exact summation 
(lines) and Monte Carlo simulation (points with error bars)
 for various aggregation types, for $w_0=0.1,\ a=0.3,\ \xi_0=1/3.5$.
Also shown is $w_0^n$, for comparison.}}
\label{f.wn}
\end{figure}

Hence it appears that a drastic change in the group behaviour is
expected to occur when the average degree of mutual influence  
represented by (\ref{e.kal},\ref{e.Kappro}) starts to compensate for 
the statistical ``insignificance" of an individual (observe the 
compensation in the exponent of
eq. (\ref{e.Wappro})).

We now turn to the exact solutions to see more precisely what happens. 
Since the number of partitions in the sum in (\ref{e.wt}) increases as
$2^{\frac{n(n-1)}{2}}$ we could sum exactly the combinatorial 
formula (\ref{e.wt}) to evaluate $W(n)$ only for
$n$ up to $\sim 7$. An exact algorithm based on the iterative solution 
(\ref{e.wit1},\ref{e.wit2}) allowed us 
to go up to $n$ about 11, we restricted ourselves to
$n=9$ for systematic runs. Large $n$ were  reached 
by Monte Carlo analysis using (\ref{e.mc1}-\ref{e.mc0}). 

For a check of the MC accuracy  we compare in
 Fig. \ref{f.xk0} the MC calculation with the exact 
summation for the case of homogeneous interaction 
(\ref{e.witc1},\ref{e.witc2})
(infinite correlation length $\xi$). 
Here and below the errors quoted do not 
account for statistic correlations in the data and
are only indicative for the stability of the latter. In the region of
very small probabilities the Monte Carlo data for large $n$ are seen to 
systematically underestimate the result by up to $30\%$, which may be
due to insufficient thermalization of our runs (we start with a
random bond configuration and perform 10000 thermalization 
sweeps at each $n$)
 -- otherwise the 
agreement is very good. 

We also can see on Fig. \ref{f.xk0} that the ``refined 
approximation" of section \ref{s.aap} does provide a lower
limit and is better than the mean field approximation in the region of the
minimum of $W(n)$, while the latter describes more accurately the asymptotic regime. 
Both of them, however, are rather far from the exact and MC results.
While providing qualitative insights and predicting
correctly the {\it position} of
the minimum and the asymptotic behaviour, 
the mean field approximation fails even at the semi-quantitative level: the
value of $W(n)$ near the minimum is underestimated by orders of magnitude. 

For the more realistic (finite correlation length $\xi$)
case we show in Fig. \ref{f.wn} numerical results 
(exact summation  
and Monte Carlo simulation) for distance
dependent interaction $K$ (\ref{e.kij},\ref{e.kdist}), using
$w_0=0.1$, $a=0.3$ and $\xi_0=1/3.5$  for various
types of aggregation: $\alpha=0$, 0.5, 0.6, 0.65, 0.75 and 1 (\ref{e.xi}). 
We see that even for small $\alpha$ (extensive aggregation) 
the presence of correlations
can increase $W(n)$ by a large factor. The most interesting result is,
however,  
the indication of the existence of an $\alpha_0$ much below 1,
such that for $\alpha > \alpha_0$ 
$W(n)$ does indeed develop a minimum after which it grows to 1 as suggested
by (\ref{e.nmin}) and (\ref{e.Wappro}).
The minimum is rather shallow and
can appear already at small $n$. Hence, ensembles which 
do not ``expand" fast enough with increasing number of points (i.e., for which
the size of the system measured in correlation lengths, $n/\xi$
increases only as a small power of $n$, $1-\alpha < 1-\alpha_0$) are
intrinsically unstable under induced behaviour. 

As a side remark, 
we notice again the large difference to the mean field approximation,
especially in the interesting
turn over region -- compare  Figs. \ref{f.mfa} and \ref{f.wn}. 
This points to the benefit of having 
 exact solutions and algorithms allowing faithful numerical
analyses.

Consider now an ensemble whose spatial organization can vary, for fixed $n$, 
then we can ask:
\par\medskip  

\no {\bf Question 2:} {\it Assuming a constant interaction scale, how does 
$W$ behave if the density of the points varies?} 
\par\medskip  

\begin{figure}[tb]
\vspace{6.5cm}
\includegraphics{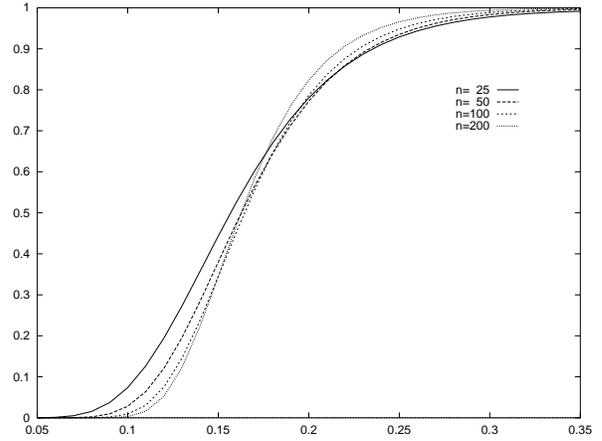}
\caption{{\it The dependence of $W$ on $K$  
for the homogeneous interaction case (infinite correlation length)
(\ref{e.witc1},\ref{e.witc2}) for $w_0=0.2$ and 
$n=25,\ 50,\ 100$
and $200$. $W$ is plotted here vs $\rho$ (the density)  
using formula (\ref{e.kd0}) to bundle the curves. 
}}
\label{f.wd0}
\end{figure}

\no Roughly, this means that the strength of the correlation varies.
In Fig. \ref{f.wd0} we show the dependence of $W$ on the
density $\rho$ (i.e., on $K$) 
for the homogeneous interaction case, 
$\xi \rightarrow \infty$
(\ref{e.witc1},\ref{e.witc2}) (exact summation) for $n=25,\ 50, 100$
and $200$ using the {\it ad hoc} rule:
\be
K(n)=\rho (20/n)^{0.78}
\label{e.kd0}
\ee
to  bundle the data and allow comparison of various $n$.

\begin{figure}[tb]
\vspace{6.5cm}
\includegraphics{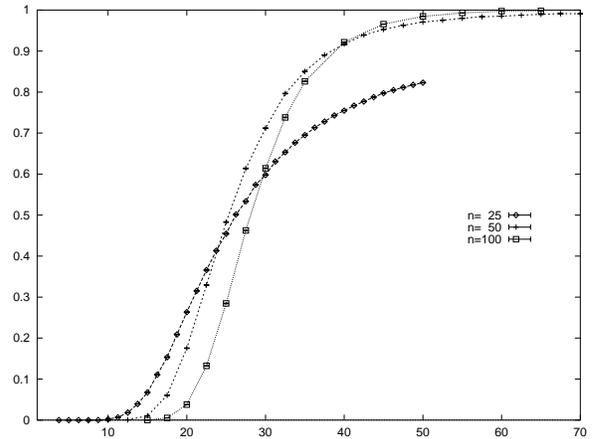}
\caption{{\it $W$ vs $\rho$ (the density) for the general case from the 
 Monte Carlo simulation for $n=25$, $n=50$ and $n=100$ and 
$w_0=0.1$, $a=0.3$ and $\xi_0=1/3.5$. 
}}
\label{f.wd}
\end{figure}

 For the general case (finite $\xi$) we show in Fig. \ref{f.wd}  
  $W$ from the Monte Carlo simulation
for three values of $n$ as function of the density $\rho$, 
where we take 
\be
\xi(n,\rho) = \xi_0 \rho.
\label{e.dens}
\ee 
\no We see that fluctuations of the density can easily destabilize the 
ensemble if the latter is near some ``critical" density, $\rho_c(n)$. 
For large $n$ the critical
fluctuations  appear to be $\propto const/\sqrt{n}$.

Also other ways of introducing a scale or for posing the stability question
 can be imagined. 
In the above discussion the parameters have been chosen more or less 
arbitrarily.
Of course, the explicit results depend on the particular problem: the form
of the function $K(d)$, the  spatial arrangement, 
the  aggregation form etc. It seems, however,
that we see here a generic feature of induced behaviour, namely the
capability to produce collective effects and that we are able in the frame
of this probabilistic model to provide a  quantitative analysis of this capability.

\end{document}